\documentclass[journal]{IEEEtran}

\usepackage[usenames, dvipsnames]{color}
\usepackage{multirow}
\usepackage{booktabs}
\usepackage{amsmath}
\usepackage{amsfonts}
\usepackage{multirow}
\usepackage{epsfig}
\usepackage{psfrag}
\usepackage{subfigure}
\usepackage{pifont}
\usepackage{cite}
\usepackage{url}
\newcounter{numberlistc}
\usepackage{algorithm}
\usepackage{algorithmic}
\usepackage{graphicx}
\usepackage[usenames]{color}
\usepackage{amssymb,amsmath}
\usepackage{amsthm}
\usepackage{subfigure}

\newcounter{itemlistc}

\usepackage{setspace}

\usepackage[font=small, labelfont=bf]{caption}

\begin{document}

\title{GPU Ising Computing for Combinatorial Optimization
  Problems in VLSI Physical Design}

\date{}

\author{
  \indent Chase Cook~\IEEEmembership{Student Member,~IEEE}, Hengyang Zhao~\IEEEmembership{Student Member,~IEEE}, Takashi Sato~\IEEEmembership{Member, ~IEEE}, Masayuki Hiromoto~\IEEEmembership{Member, ~IEEE}, Sheldon
  X.-D. Tan~\IEEEmembership{Senior Member,~IEEE}

~\thanks{
\indent Chase Cook, Hengyang Zhao, and Sheldon X.-D Tan are with the Department of
Electrical and Computer Engineering,  University of California,
Riverside, CA 92521.
\newline \indent Takashi Sato and Masayuki Hiromoto are with Department of
Communication and Computer Engineering, Graduate School of
Informatics, Kyoto University, Kyoto, Japan. 
}
}






\maketitle

\begin{abstract}
  In VLSI physical design, many algorithms require the solution of
  difficult combinatorial optimization problems such as max/min-cut,
  max-flow problems etc. Due to the vast number of elements typically
  found in this problem domain, these problems are computationally
  intractable leading to the use of approximate solutions.  In this
  work, we explore the Ising spin glass model as a solution
  methodology for hard combinatorial optimization problems using the
  general purpose GPU (GPGPU).  The Ising model is a mathematical
  model of ferromagnetism in statistical mechanics. Ising computing
  finds a minimum energy state for the Ising model which essentially
  corresponds to the expected optimal solution of the original
  problem. Many combinatorial optimization problems can be mapped into
  the Ising model. In our work, we focus on the max-cut problem as it
  is relevant to many VLSI physical design problems.  Our method is
  inspired by the observation that Ising annealing process is very
  amenable to fine-grain massive parallel GPU computing. We will
  illustrate how the natural randomness of GPU thread scheduling can
  be exploited during the annealing process to create random update
  patterns and allow better GPU resource utilization. Furthermore, the
  proposed GPU-based Ising computing can handle any general Ising
  graph with arbitrary connections, which was shown to be difficult
  for existing FPGA and other hardware based implementation
  methods. Numerical results show that the proposed GPU Ising max-cut
  solver can deliver more than 2000X speedup over the CPU version of
  the algorithm on some large examples, which shows huge performance
  improvement for addressing many hard optimization algorithms for
  practical VLSI physical design.
\end{abstract}

\section{Introduction}
\label{sec:intro}
   
There are many hard combinatorial optimization problems such as
max-flow, max-cut, graph partitioning, satisfiability, and tree based
problems, which are important for many scientific and engineering
applications. With respect to VLSI physical designs, these problems
translate to finding optimal solutions for cell placement, wire
routing, logic minimization, via minimization, and many others.  The
vast complexity of modern integrated circuits (ICs), some having
millions or even billions of integrated devices, means that these
problems are almost always computationally intractable and require
heuristic and analytical methods to find approximate solutions.  It is
well-known that traditional von Neumann based computing can not
deterministically find polynomial time solutions to these hard
problems~\cite{Papadimitriou:Book'98}.
 

To mitigate this problem, a new computing paradigm utilizing the {\it
  Ising spin glass model} or {\it Ising model} has been
proposed~\cite{HNishimori:book'01}. The Ising model is a mathematical
model describing interactions between magnetic spins in a 2D
lattices~\cite{JMydosh:book'93}. The model consists of {\it spins},
each taking one of two values \{+1,-1\} (to represent up and
down states of a spin along a preferred axis) and are generally
arranged in a 2D lattice. The spin's value is determined so that its
energy is minimized based on interactions with its neighbor
spins. Such local spin updates will lead to the ground state (globally
lowest energy configuration) of the Ising model. It was shown that
many computationally intractable problems (such as those in class NP
complete or NP hard) can be converted into Ising
models~\cite{ALucas:FIP'14}. Some natural processes, such as quantum
annealing process, were proposed as an effective way for finding such
a ground
state~\cite{MJohnson:Nature'11,BSergio:NatPhys'14}. D-Wave~\cite{dwave}
is one such quantum annealing (also called adiabatic quantum
computation) solver based on the Ising model and it shows $10^8$
speedup over simulated annealing on the weak-string cluster pair
problem\cite{VDenchev:PRX'16}.  However, existing quantum annealing
requires close to absolute zero temperature operating on
superconductive devices, which are very complicated and expensive.

While quantum computing has yet to reach maturity, there exists a
number of other hardware-based annealing solutions which have been
proposed to exploit the highly parallel nature of the annealing
process in the Ising model.  In~\cite{MYamaoka:IJSSC'16}, a novel CMOS
based annealing solver was proposed in which an SRAM cell is used to
represent each spin and thermal annealing process was emulated to find
the ground state. In \cite{HGyoten:IEICE'18,CYoshimura:IJNC'17}, the
FPGA-based Ising computing solver has been proposed to implement the
simulated annealing process. However, those hardware based Ising model
annealing solvers suffer several problems. First, the Ising model for
many practical problems can lead to very large connections among Ising
spins or cells.  Furthermore, embedding those connections into the
2-dimensional fixed degree spin arrays in VLSI chips is not a
trivial problem. Doing so  requires mitigation techniques such as cell
cloning and splitting using graph minor embedding (another NP-hard problem) as proposed
in~\cite{HGyoten:IEICE'18,CYoshimura:IJNC'17,Gyoten:ICCAD'18}. Second,
ASIC implementations are not flexible and can only handle a specific
problem due to the fixed topology among spins, and FPGA
implementations require architectural redesign, and thus recompilation, for each different
problem. Third, one has to design hardware for the random number
generator for each spin cell and simulate the temperature changes,
which occupies significant chip area resulting in scalability
degradation.

Based on the above observations and the highly parallel nature of the
Ising model, in this work, we explore the General Purpose Graphics
Processing Unit (GPGPU or simply GPU) as the Ising model annealing computing
platform. The GPU is a general computing platform, which can provide
much more flexibility over VLSI hardware based annealing solutions as
a GPU can be programmed in a more general way, enabling it to handle
any problem that can be mapped to the Ising model. That is, it is not
restricted by the topology or complex connections that some problems
may have. At the same time, it provides massive parallelisms compared
to existing CPUs. The GPU is an architecture that utilizes large
amounts of compute cores to achieve high throughput performance. This
allows for very good performance when computing algorithms that are
amenable to parallel computation while also having very large data sets
which can occupy the computational resources of the
GPU\cite{cuda,NvidiaKepler}. The problem sizes in the physical
design domain can easily accomplish this and heuristic methods can
solve the Ising model in a parallel manner which makes the GPU ideal
for this application. We remark that extensive work for Ising
computing on GPUs have been proposed
already~\cite{BBlock:CPC'10,LBarash:CPC'17,MWeigel:JCP'12}, however;
they still focus on physics problems which assume a nearest neighbor
model only. This model, is highly amenable to the GPU computing as it
is easily load balanced across threads but is not general enough to
handle problems such as max-cut without extensive preprocessing. Furthermore, many GPU-based methods
use a checkerboard update scheme, but this is still only practical for
the nearest neighbor model without using complicated graph embedding.

In this paper we propose a GPU-based Ising model solver, using a modified
simulated annealing heuristic, that can handle any general problem.
We focus on the max-cut problem as it is relevant to many VLSI
physical design problems.  We show that Ising computing by the
simulated annealing process is very amenable to fine-grain GPU-based
parallel computing. We further propose an update method that utilizes
the GPU scheduler to achieve a random update pattern enabling independent parallel spin updates. This
allows us to maximize thread utilization while also avoiding
sequential and deterministic update patterns for a more natural
annealing process.  Furthermore, the new GPU-based Ising computing
algorithm can handle any general Ising graph with generally connected
spins, which was shown to be difficult for FPGA and other hardware
based implementation methods. Our numerical results show that the
proposed GPU Ising solver for max-cut problem can deliver more than
2000X speedup over the CPU version of the algorithm on some large
examples, which shows huge performance improvement.

\section{Ising model and Ising computing}
\label{sec:ising_algorithm}
   
\subsection{Ising model overview}
\label{sec:ising}

\begin{figure}[!ht]
  \centering
  \includegraphics[width=.60\columnwidth]{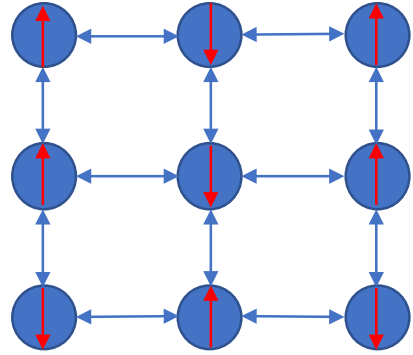}  
  \caption{The 2D nearest neighbor Ising model.}
  \label{fig:ising_model}
\end{figure}

The Ising model consists of a set of spins interconnected with each
other by a weighted edge. For the general Ising model, spin
connections can take on any topology. One of the connection
topologies is the 2D lattice, referred to as the nearest neighbor
model, shown in Fig.~\ref{fig:ising_model}, which describes the
ferromagnetic interactions between so-called spin glasses. Many
computationally intractable problems can be mapped to this Ising
model. It was shown that finding the ground state in the 2D lattice Ising
model is an NP-hard problem~\cite{FBarahona:JPMATH'82}. However, it
has certain characteristics that make it more amenable to the
annealing process as each local update results in energy minimization and
spin glass updates can be performed in a highly parallel manner.

 
Specifically, each spin $\sigma_i$, has two discrete spin values
$\sigma_i \in \{-1,1\}$ and some interaction with adjacent spins in the form of a weighted edge. Then the
local energy or Hamiltonian of the spin is described
by~\eqref{eq:ising_model_local}:
\begin{equation}
  {\cal H}_{i}(\sigma_i) =- \sum\limits_{j}^{} J_{i,j} \sigma_{i} \sigma_{j} - h_i \sigma_i
 \label{eq:ising_model_local}
\end{equation}

In this equation, $J_{i,j}$ is the interaction weight between
$\sigma_i$ and $\sigma_j$, and $h_i$ is a bias or external force
acting on $\sigma_i$. 

{\bf Local spin update:} by finding the minimum value of
${\cal H}_{i}(\sigma_i)$, we can determine the local spin value
$\sigma_i$. Specifically \eqref{eq:ising_model_local} can be written
as
\begin{equation}
{\cal H}_{i}(\sigma_i) = \left( -\sum\limits_{j}^{} J_{i,j} \sigma_{j} -
h_i \right) \sigma_i = - S \times \sigma_i
 \label{eq:ising_model_local_1}
\end{equation}
From \eqref{eq:ising_model_local_1}, we can see that $\sigma_i$ can be

determined just from the sign of the $S$ value. If $S > 0$,
$\sigma_i = 1$, otherwise, $\sigma_i = -1$. If $S = 0$, it can take
any value of $\{-1,1\}$. This is called a{\bf local spin update or
    update}, in Ising computing.  We note that such an update for
obtaining the minimum value of ${\cal H}_{i}(\sigma_i)$ only depends
on its neighbors. By ensuring that spin updates are not
  correlated, then all the spin updates can be done {\it
    independently} and thus {\it in
    parallel}~\cite{MYamaoka:IJSSC'16,HGyoten:IEICE'18,CYoshimura:IJNC'17,BBlock:CPC'10,LBarash:CPC'17}.
  Then the global energy of the whole Ising model is given by the
  following \eqref{eq:ising_model_global},
\begin{equation}
  {\cal H} (\sigma_1, \sigma_2, ..., \sigma_n) = -\sum\limits_{\langle i,j\rangle}^{} J_{i,j} \sigma_{i} \sigma_{j} - \sum\limits_{i} h_i \sigma_i
\label{eq:ising_model_global}
\end{equation}
Note that $\langle i,j\rangle$ indicates the combination
of all spin interactions. In general, we refer the problem
of finding the minimum energy of the Ising Model, or equivalently the
ground state of Ising Hamiltonian, as the Ising problem.  It can be
shown that the Ising problem shown is equivalent to the problem of quadratic
unconstrained Boolean optimization (QUBO)
\cite{Boros:2007:LSH:1231244.1231283}.  It was shown that many computationally
intractable problems (such as those in class NP-complete or NP-hard)
can be converted into Ising models~\cite{ALucas:FIP'14}. 
 
\begin{figure}[!ht]
  \centering
  \includegraphics[width=0.60\columnwidth]{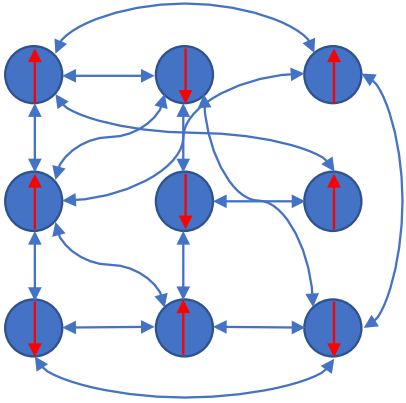}
  \caption{An example of a generally connected Ising model.}
  \label{fig:ising_model_gen}
\end{figure}

Previous methods have focused on solving the nearest neighbor Ising
model~\cite{MYamaoka:IJSSC'16,HGyoten:IEICE'18,CYoshimura:IJNC'17}. However,
this model has the drawback of not being able to handle any general
problem which may have arbitrary and complex connections. Therefore,
in this work, we assume that a spin glass's connections, or edges, are
able to connect to any other spin glass in the model, an example of
which is shown in Fig.~\ref{fig:ising_model_gen}. Using this more
general model removes the nearest neighbor restriction on the Ising
model and allows us to handle more complex problems.

From the Hamiltonians presented in this section, we can see that if we
minimize the local energy of each spin, we also minimize the global
energy of the entire system which leads to the ground state of the
model. We can then map a combinatorial optimization problem to this
Ising model such that the ground state of the Ising model corresponds
to the global solution of the corresponding optimization problem. In
this way, finding the local energy of each spin, which leads to
finding the ground state of the Ising model, is the same as finding
the optimal solution to the problem in question.

\subsection{Annealing method for Ising model solution}
\label{sec:algorithm}

\begin{figure}[ht!]
  \centering
  \includegraphics[width=\columnwidth]{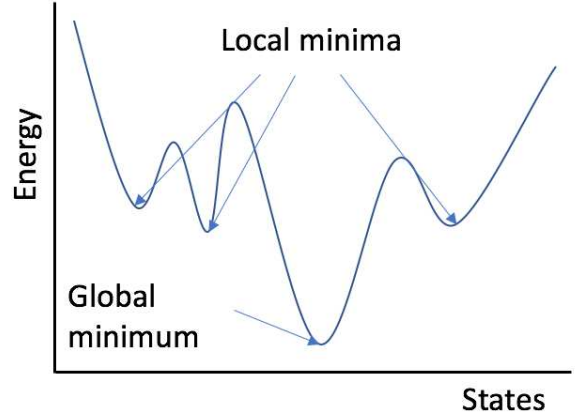}
  \caption{Depiction of the local minima and global minimum in the energy minimization problem.}
  \label{fig:energy}
\end{figure}

A classical and well known heuristic for combinatorial optimization is
Simulated Annealing (SA)~\cite{Kirkpatrick:Science'83}. This heuristic
mimics the behavior of thermal annealing, found in
metallurgy. Essentially, it works by setting the environment to a high
``temperature,'' giving the model high energy and allowing for higher
probability of changing states, and then gradually decreases the
temperature as the simulation runs. More precisely, it iteratively
calculates and evaluates the global solution quality of neighbor
states of a model and probabilistically allows the acceptance of
a new state even if its solution quality is worse than the previous
state by utilizing the Metropolis criteria. The probability of
accepting a worse state is dependent on the temperature of the system
which gradually decays over time. This allows the heuristic to avoid
local minima as depicted in Fig.~\ref{fig:energy}.


In this proposed work, we use a simplified/modified Metropolis
  annealing algorithm to find the ground state as shown in
  Algorithm~\ref{alg:annealing} that better exploits the features of
  the Ising model while also allowing us to avoid local minimas.  In
  our proposed method, we allow each spin update to minimize its own
  local energy and do not compute a global solution quality (which
  would add a large computational penalty). In order to avoid local
  minimas we add energy to our Ising model by utilizing random flip
  probabilities that decay over time.


\begin{algorithm}
  \caption{Modified Ising annealing algorithm}
  \label{alg:annealing}
  \begin{algorithmic}[1]
    \STATE {input: ($M$, $N$, ${\bf S}$)}
    \STATE{initialize all $\sigma_i$ in ${\bf S}$}
    \FOR{sweep-id \textbf{in} \{1, 2, \ldots, $M$\}}
      \FOR{$\sigma_i$ \textbf{in} ${\bf S}$}
      \STATE{$\sigma_i \leftarrow \mathrm{argmin}(H(\sigma_i))$  based
      on  \eqref{eq:ising_model_local_1}}
      \ENDFOR
      \STATE{randomly choose and flip $N$ spin glasses in ${\bf S}$}
      \STATE{decrease $N$}
    \ENDFOR  
  \end{algorithmic}
\end{algorithm}

In algorithm~\ref{alg:annealing}, $M$ is the maximum number of sweeps,
$N$ is the number of spins to randomly flip, and ${\bf S}$ is the set
of all spin glasses. Once the spin glasses are initialized, all spin
glasses are updated iteratively to propagate the interactions between
each spin (a process we will call a ``sweep'').  When a spin glass
  updates based on \eqref{eq:ising_model_local_1}, it computes and
  chooses the spin value that will minimize its local Hamiltonian.  At
the end of a sweep, $N$ glasses are randomly flipped and $N$ is
decreased according to an annealing schedule. After this, the process
is repeated for $M$ sweeps or until convergence is achieved.

Following this process, the local energy of each spin glass, and thus
the global energy of the model, is gradually decreased. Furthermore, we
avoid the local minima by introducing energy to the model by using
uncorrelated random flips which slowly decrease over time.

We want to remark that the modified Ising annealing process can be
viewed as a simplified Metropolis Monte Carlo method as we essentially
accept the positive energy changes of a spin flip based on a
temperature dependent probability~\cite{Metropolis:1953in}. As a
result, we should have the convergence properties of the Metropolis
method~\cite{DLandau:book'05}, which means that the objective function
will be minimized in a statistical way for a sufficient time. We
notice that similar Ising annealing processes have been used for FPGA
and CMOS based Ising
computing~\cite{MYamaoka:IJSSC'16,HGyoten:IEICE'18,Gyoten:ICCAD'18}.



\section{Problem description}
\label{sec:max-cut}
  
The Ising model and the method introduced in this paper can be applied
to many NP class problems, however, we use the max-cut problem as a
practical example. The max-cut problem, in practice, can help find
solutions to several EDA and VLSI design problems. For example, the
general via minimization problem, the act of assigning wire segments
to layers of metallization such that the number of vias is minimized,
can be modeled as a max-cut
problem~\cite{FBarahona:OR'88,FBarahona:TCAS'90,JCho:TComp'98}.

The max-cut problem is highly amenable to the Ising model, making
  it an ideal candidate to introduce the proposed
  methodology. Furthermore, we also note that the Ising spin model and
  max-cut have been used as a solution technique for the
  via-minimization problem in the past\cite{FBarahona:OR'88}.  

The via minimization problem is relevant to the VLSI domain as
  vias are a source of manufacturing defects in ICs which can cause
  reductions in yield. Therefore, it is desirable to minimize the
  number of required vias in an IC~\cite{NHarrison:DFTVS'01}.  

The 2-layer via minimization problem can be formulated by
  constructing a conflict graph from a transient routing of
  interconnects where each interconnect is represented by a node in
  the graph~\cite{FBarahona:OR'88,FBarahona:TCAS'90,JCho:TComp'98}. Interconnects with
  routes that cross are considered in conflict (as they cannot be
  placed in the same layer of metal) and are connected by a ``conflict
  edge'' in the graph. Wire segments that are not in conflict are
  connected by a ``continuation edge'' . For every cluster of nodes
  connected by a conflict edge, we can collapse the nodes into a
  single representative node such that only continuation edges are
  left in the graph. A cut of this graph, or a grouping of nodes, then
  corresponds to assigning wire segments to a layer of metal. We then
  assign two initial weights to each edge, $a_{ij}$ and $b_{ij}$,
  which respectively correspond to the number of vias required if the
  edge is cut and the number of vias required if the edge is not
  cut. A combined weight is then calculated as $w_{ij} = b_{ij} -
  a_{ij}$. At this point, the maximum cut of this graph corresponds to
  minimizing the number of vias.
 
The max-cut problem is defined as partitioning a graph into two subsets
$S$ and $\bar{S}$ such that the weighted edges between the vertices of one
subset and the other subset are maximized. This is mathematically
formulated by~\eqref{eq:max-cut} assuming a graph $G = (V,E)$ has a variable
$x_i$ assigned to each vertex:

\begin{equation}
  \begin{split}
  &\max \frac{1}{2} \sum\limits_{i,j\in V,i<j} w_{i,j}(1-x_i x_j) \\
  &\mathrm{s.t.} \quad  x_i \in \{1,-1\}\\
  \end{split}
  \label{eq:max-cut}
\end{equation}

In this equation, $V$ is the set of vertices in the graph $G$, $w_{i,j}$ is
the edge weights in $E$ between the $i$th and $j$th elements in $V$, and $x_i$ is an indication of
which subset the vertex belongs to and can take the values $\{-1,1\}$.

Intuitively, looking as the Ising spin glass problem in
\eqref{eq:ising_model_global}, we can see how the max-cut problem
should map to the Ising model by associating the spin of a spin glass
$\sigma_i$ with a subset of the graph in the max-cut problem. That is,
we can say that if a spin is $1$ then the spin glass is in $S$ and if a
spin is $-1$ then it is in $\bar{S}$, which is analogous to
$x_i$. Furthermore, the weights between vertices $w_{i,j}$ is the same
as the interaction weights between spin glasses $J_{i,j}$ and, in this
case, there is no bias or external force so the $h$ term in the Ising
model is simply zero. The global energy minimization of the Ising
model for the max-cut problem is shown below
in~\eqref{eq:ising_max-cut}:

\begin{equation}
  \centering
  {\cal H} = - \sum \limits_{\langle i,j\rangle} J_{i,j}\sigma_i\sigma_j
  \label{eq:ising_max-cut}
\end{equation}

Once mapped to the Ising model, the max-cut problem can then be solved
by finding the ground state of the model using the methods proposed in
this paper. While there are other ways to solve this problem, the
method we propose is highly amenable to parallel computation and large
problem sets have great performance when implemented on the GPU, thus,
giving our method an advantage in scalability.

\section{GPU Implementation}
\label{sec:gpu_impl}
  
  
\subsection{GPU Architecture}
The general purpose GPU is an architecture designed for highly
parallel workloads which is leveraged by Nvidia's CUDA, Compute
Unified Device Architecture, programming model~\cite{cuda}.  GPU's
offer massive parallelism favoring throughput oriented computing in
contrast to traditional CPUs which focus primarily on latency
optimized computation.  This gives the GPU an advantage in scalability
so long as a problem does not contain large amounts of sequential
operations. The Nvidia GPU architecture is comprised of several
Symmetric Multiprocessors (SMs), each containing a number of ``CUDA''
cores, and a very large amount of DRAM global
memory~\cite{NvidiaKepler}. The Kepler architecture based Tesla K40c
GPU, for example, has 15 SMs for a total of 3072 CUDA cores (192 cores
per SM), and 12GB DRAM global memory. Additionally, each SM has
several special function units, shared memory, and its own cache.

\begin{figure}
  \centering
    \resizebox{.4\textwidth}{!}{\input{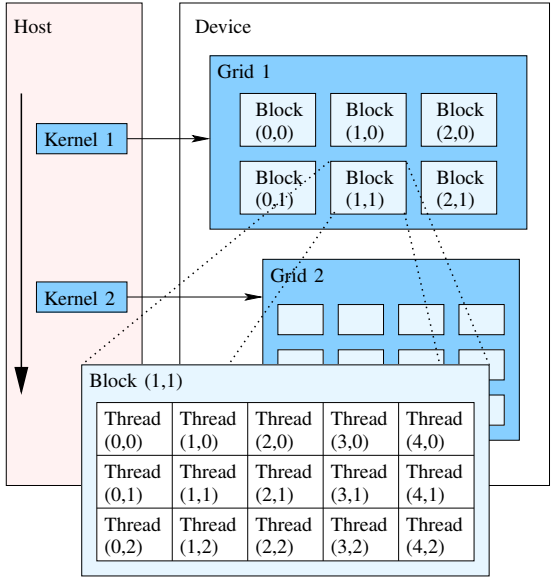}}
    \caption{The Nvidia CUDA programming model showing the Host (CPU) and Device (GPU) and the relation between threads, blocks, and grids.}
    \label{fig:cuda_model}
    
\end{figure}

The CUDA programming model, shown in Fig.~\ref{fig:cuda_model},
extends the C language adding support for thread and memory allocation
and also the essential functions for driving the
GPU~\cite{CUDA_C_Programming_Guide_2018}. The model makes a
distinction between the host and device or the CPU and GPU
respectively. The model uses an offloading methodology in which the
host can launch a device kernel (the actual GPU program) and also
prepare the device for the coming computation, e.g., the host will
create the thread organization, allocate memory, and copy data to the
device. In practice, a programmer must launch many threads which will be
used to execute the GPU kernel. Thread organization is therefore
extremely important in GPU programming. Threads are organized into
blocks which are organized into grids. Each block of threads also has
its own shared memory which is accessible to all the threads in that
block. Additionally, the threads in the block can also access a global
memory on the GPU which is available to all threads across all blocks.

The GPU fundamentally focuses on throughput over speed. This
throughput is achieved through the massive compute resources able to
be run in parallel. Because of this, it is important to realize that
the GPU is not meant for small data sets or extremely complicated
operations that may be better suited for a powerful CPU. Instead, the
GPU is meant to execute relatively simple instructions on massive data
in parallel that can occupy the GPU resources for an extended period
of time. This computing paradigm gives the GPU a huge advantage in
scalability so long as it has sufficient hardware resources for the
problem being solved. So long as this is the case, and there is no
significant sequential operations in the GPU kernel, the GPU
computation time for a problem will not grow significantly.


\subsection{Ising model implementation}

The GPU, while lacking the massive scaling of a quantum computer,
has much larger scaling capabilities than CPUs, allowing it to handle
very large problem sizes. Indeed, smaller problems that are unable to
fully utilize the GPU resources may achieve worse performance
than a CPU.

In order to ensure the best utilization of GPU resources, it is
necessary to devise a spin glass update scheme more amenable to
parallel computation. Algorithm~\ref{alg:annealing} relies on
sequential updates to propagate the interactions between the spin
glasses. However, this would be highly inefficient on the GPU as it
would mean each thread would have to wait for previous threads to
update.

In previous works that have addressed the nearest neighbor Ising
model, a checkerboard update scheme is implemented which allows for
many spin glasses to be updated in
parallel~\cite{MWeigel:JCP'12}. While the spin glass updates are
independent of their neighbors, they are not truly independent since
the update pattern is deterministic and can introduce autocorrelation
between spin updates but this autocorrelation generally does not
affect the global balance of the model~\cite{MWeigel:JCP'12}. For the
problem addressed in this work, however; interactions are not confined
to nearest neighbors nor are they restricted to regular patterns. This
results in very complex interactions in which spins can be dependent
on many other spins across the entire model. Consequently, a different
update scheme must be developed for such a general solver to ensure the independence of parallel updates.

To address the above mentioned issues, we modify the original
algorithm in algorithm~\ref{alg:annealing}. Firstly, we assign each
thread to a single spin glass, and make it responsible for updating
that glass. One may notice that since each spin glass may have a
different number of neighbors, then the threads will not be perfectly
load balanced. However, the alternative is to use graph minor
embedding, another NP-hard problem~\cite{JCai:arXiv'14}, to create
clone nodes such that every thread will have an equal number of
updates~\cite{HGyoten:IEICE'18}. However, this means that we need to
have the CPU do some intensive pre-processing on the model, and it
also means that after each update sweep, a reduction must be performed
on each spin glass's clones so that the true spin value can be
determined. For these reasons, it is much better to allow for some
load imbalance and suffer some computational penalty on the GPU,
instead of increasing the complexity of the algorithm. Furthermore,
while each update sweep is synchronized, we do not synchronize the
updates of each spin which makes them uncorrelated. In practice, this means that threads will update
their assigned spin glass as soon as they are scheduled and will use
whatever spin status their respective neighbors have at the time of
data access. Therefore, there is no guarantee that a spin's neighbors
will contain the spin value from the current sweep or from the previous sweep. This naturally
propagates the updates of each spin glass in a non-deterministic
pattern which ensures that each spin update can be done in independently of its neighbor and in parallel.



Another major change to the algorithm is the implementation of the
random flips. Instead of randomly selecting a number of spin glasses
to be flipped at the end of an update sweep, we let each thread independently decide
if it should flip or not. A global variable, visible to all threads,
gives the flip probability in the form of a floating point number
between 0.0 and 1.0. Each thread then independently generates a random number between
0.0 and 1.0, generated by the CUDA cuRAND library for efficient random
number generation, and flips if the number is below the global flip
probability. The cuRAND library used for the random number
generations allows us to create a random number generator for each
thread at the very beginning of the program. Each thread is then
responsible for calling its own generator when it requires a random
number. The advantage of using the cuRAND library is that the random
numbers can be generated completely in parallel and independent of each other by allowing the GPU
threads to do their own random number generation work. This also is
amenable to our asynchronous update scheme since each thread is
responsible for its own flipping and can decide to flip or not as soon
as it finishes its update without waiting for other threads which may
still be updating. Consequently, a thread may not only update using an
already updated neighbor, it may actually update using a neighbor that
has been randomly flipped also. The flip probability is then reduced
as update sweeps are completed.

\begin{algorithm}
  \caption{GPU Simulated Annealing method for Ising model}
  \label{alg:annealing_gpu}
  \begin{algorithmic}
    \STATE{ input: ($F_p, {\bf S}$)}
    \STATE{initialize ALL $\sigma_i$ in ${\bf S}$}
    \WHILE{$F_p > 0$}
    \FOR{all $\sigma_i \in \bf S$ \textbf{in parallel}}
      \STATE{$\sigma_i \leftarrow \mathrm{argmin}(H(\sigma_i))$}
      \STATE{flip $\sigma_i$ with probability $F_p$}
     \ENDFOR
     \STATE{reduce $F_p$}
     \ENDWHILE
  \end{algorithmic}
\end{algorithm}

In algorithm~\ref{alg:annealing_gpu}, the parallel simulated annealing
solver for the Ising model is presented. While mostly similar to the
algorithm~\ref{alg:annealing}, There are some differences. We replace
the number of random flips input with a variable $F_p$ which
represents the flip probability mentioned above. Next, we have each
thread generate a random number, using the cuRAND library, between 0.0
and 1.0 and compare this to $F_p$. If it is less, then the thread will
flip the spin value. While these changes are subtle, the effect is
large as it allows the parallel computation of an entire update sweep.

\section{Experimental Results}
\label{sec:results}
 
In this section, we present the experimental results showing both the
accuracy and speed of our parallel GPU-based Ising model solver for
the max-cut problem. The CPU-based solution is done using a Linux
server with 2 Xeon E5-2698v2 2.3~GHz processors, each having 8 cores (2 threads per
core for a total of 32 threads), and 72~GB of memory. On the same
server, we also implement the GPU-based solver using the Nvidia Tesla
K40c GPU which has 3072 CUDA cores and 12~GB of memory. Test problems
from the G-set benchmark~\cite{gset} are used for testing as well as
some custom made problems to show very large cases. The problems' edge
counts are used to represent their size as the size of each problem is
dominated by the number of edges.

\subsection{Accuracy study}
\label{sec:accuracy}

To test the accuracy of the method presented in this paper, we compare
the max-cut value our method generates with that of the best known
solution in the G-set benchmark. In addition to the G-set benchmark
comparison, we generated several custom graphs and compared the
solution quality of our GPU-based method against IBM's CPLEX
mathematical programming solver~\cite{ibm_cplex}, a state-of-the-art linear programming solver employed to solve combinitorial optmization problems. CPLEX solutions were
implemented using a server with a 2.1~GHz Xeon Broadwell processor
with 36  threads (18 cores with 2 threads per core) and 128~GB of
memory.

\begin{table}[h!]

  \caption{Accuracy comparison of the GPU max-cut value against the best known cut values for the G-set benchmark.}
  \centering
  \begin{tabular}{|c|c|c|}
    \hline Graph & Best known cut & GPU cut (\%accuracy) \\
    \hline\hline G13 & 580 & 522(90.0\%) \\
    G34 & 1372 & 1191(86.8\%) \\
    G19 & 903 & 844(93.5\%) \\
    G21 & 931 & 880(94.6\%) \\
    G20 & 941 & 880(93.6\%) \\
    G18 & 988 & 938(94.9\%) \\
    G51 & 3846 & 3754(97.6\%) \\
    G53 & 3846 & 3756(97.7\%) \\
    G54 & 3846 & 3756(97.7\%) \\
    G50 & 5880 & 5803(98.7\%) \\
    G47 & 6656 & 6619(99.4\%) \\
    G40 & 2387 & 2267(95.0\%) \\
    G39 & 2395 & 2269(94.7\%) \\
    G42 & 2469 & 2325(94.2\%) \\
    G41 & 2398 & 2284(95.2\%) \\
    G9 & 2048 & 2004(97.9\%) \\
    G31 & 3288 & 3227(98.2\%) \\ \hline
\end{tabular}

    \label{tb:accuracy}
\end{table}

\begin{table}[h!]
  
  \caption{Accuracy comparison of the GPU max-cut value against the cut values obtained by CPLEX.}
  \centering  
  \begin{tabular}{|c|c|c|}
    \hline \# edges & CPLEX cut & GPU cut (\%accuracy) \\
    \hline\hline 9999 & 9473 & 8884 (93.78\%) \\
    14999 & 13357 & 12776(95.65\%) \\
    24998 & 20206 & 19981(98.88\%) \\
    49995 & 35248 & 36228(100.29\%) \\
    39998 & 33605 & 32914(97.94\%) \\
    59997 & 46371 & 46510(100.29\%) \\
    99995 & 70566 & 72009(102.04\%) \\
    199990 & 128448 & 131930(102.71\%) \\
    249995 & 176556 & 179391(101.60\%) \\
    374993 & 248505 & 255078(102.64\%) \\
    626988 & 392912 & 400540(101.94\%) \\
    1249975 & 741709 & 751050(101.25\%) \\ \hline
\end{tabular}

    \label{tb:accuracy2}
\end{table}

 For the results in Table~\ref{tb:accuracy} and
  Table~\ref{tb:accuracy2}, cut values were obtained by running the
  GPU solver 10 times per graph and taking the average solution
  quality. Each time the solver was run we used 1000 annealing steps
  which is the same number used to generate the performance data in
  the following performance study section. Additionally, the number of
  random flips for each solution is set to 200 which decays linearly
  each sweep by a factor of $0.01$. The cut values generated by CPLEX
were obtained by running the CPLEX solver for each graph case. The
solver time for CPLEX was capped at two days with most cases using all
of the allocated time. We also note that we don't compare the solution
time to our solver since the CPLEX solution time is on the order of
days while the GPU Ising solver's solution time is on the order of
seconds.  The {\it accuracy\%} is defined as the ratio of the cut
values of the GPU Ising solver over the best known
cut(Table~\ref{tb:accuracy}) or the CPLEX solver
cut(Table~\ref{tb:accuracy2}), i.e., the closer to 100\% the better
the {\it accuracy\%}. If {\it accuracy\%} is larger than 100\%, then
GPU Ising solver obtained a better result than the competing solver.

From the tables, we can see that the GPU
consistently performs well with almost all results above
$90\%$. Furthermore, in Table~\ref{tb:accuracy2} our GPU-based solver
is able to consistently beat CPLEX for larger cases which become too
large for CPLEX to solve in a reasonable amount of time. We also note
that in practice, the best result could be picked from a number of
GPU simulations and some simulation parameters could be tuned to achieve
better results, e.g., annealing schedule and initial flip
probability. However, we present average results of a parameter
configuration we found to be consistent across many graphs.

\begin{figure}
  \centering
  \includegraphics[width=\columnwidth]{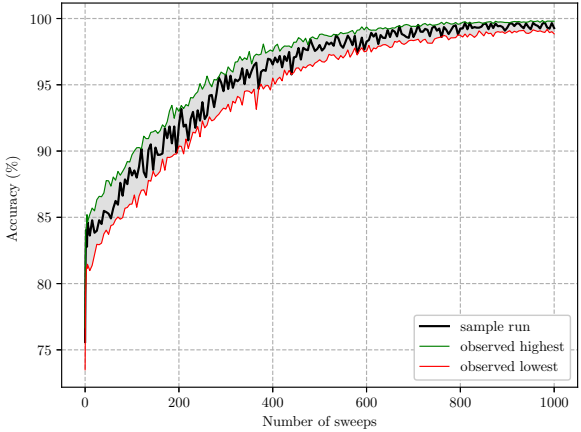}
  \caption{Convergence region of the GPU-based annealing for the Ising model on the G-set G47 problem.}
  \label{fig:converge}
\end{figure}

In Fig.~\ref{fig:converge} the region of convergence is shown for the
GPU Ising solver and was obtained by running the solver several times
for a particular problem. The red line shows the lowest observed
accuracy while the green line shows the highest observed accuracy. One
example run is also included to show the overall behavior of the
convergence. Unlike classic simulated annealing, the solver does not
converge to a single state but rather it continues to have minor
variations in energy as the solver progresses. This is because of the
utilization of the GPU scheduler as the random update pattern which
means there will be some oscillations between solution states, even
when the number of random flips is small or zero.

\subsection{Performance study}
\label{sec:speed}

Next, we look at the GPU Ising solver's performance.  The speed of the
GPU Ising solver is judged by measuring how long it takes to perform a
number of update sweeps on various sized models. We run both the
proposed GPU Ising solver and the sequential CPU implementation of the
algorithm for 1000 sweeps (which is the same number of sweeps used to
generate the accuracy results above) and compare the computation time. Because
the GPU performs best when its resources are fully utilized, and
because it is not optimized for small workloads, we expect to see
performance gain over the cpu to improve as the problem size increases. We also should not
expect a large speed-up over a CPU version for smaller problems.

\begin{figure*}
  \centering
  \includegraphics[width=\textwidth]{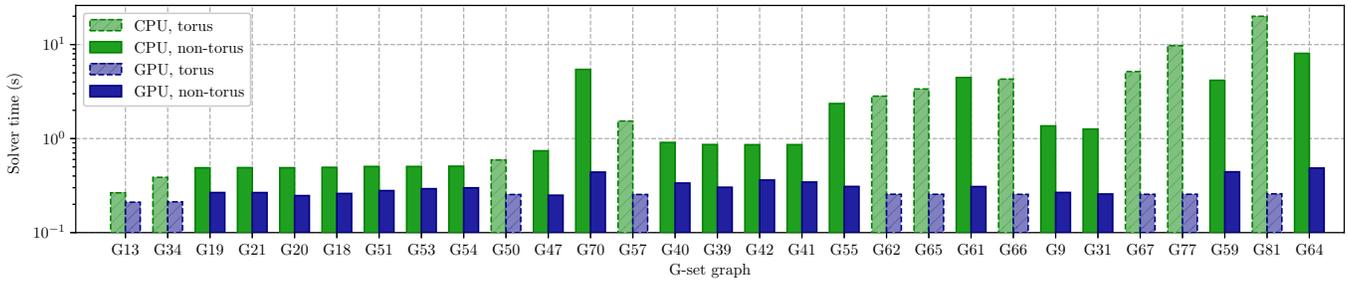}
  \caption{Speedup results of the GPU against the CPU for the G-set benchmark problems.}
  \label{fig:cpu_v_gpu_bars}
\end{figure*}

\begin{table}
    \footnotesize
\begin{tabular}{|c|c|c|c|c|c|c|}
    \hline
    graph & \#vertices & \#edges & torus & $t_\text{CPU}$ (s) & $t_\text{GPU}$ (s) & speedup \\
    \hline
    \hline
    G13 & 800   & 1600  & yes   & 0.17      & 0.11  & 1.5       \\
    G34 & 2000  & 4000  & yes   & 0.29      & 0.11  & 2.6       \\
    G19 & 800   & 4661  & no    & 0.39      & 0.17  & 2.3       \\
    G21 & 800   & 4667  & no    & 0.39      & 0.17  & 2.3       \\
    G20 & 800   & 4672  & no    & 0.39      & 0.15  & 2.6       \\
    G18 & 800   & 4694  & no    & 0.39      & 0.16  & 2.5       \\
    G51 & 1000  & 5909  & no    & 0.41      & 0.18  & 2.3       \\
    G53 & 1000  & 5914  & no    & 0.41      & 0.19  & 2.1       \\
    G54 & 1000  & 5916  & no    & 0.41      & 0.20  & 2.0       \\
    G50 & 3000  & 6000  & yes   & 0.49      & 0.15  & 3.2       \\
    G47 & 1000  & 9990  & no    & 0.64      & 0.15  & 4.3       \\
    G70 & 10000 & 9999  & no    & 5.34      & 0.34  & 15.7      \\
    G57 & 7000  & 10000 & yes   & 1.44      & 0.15  & 9.3       \\
    G40 & 2000  & 11766 & no    & 0.81      & 0.24  & 3.4       \\
    G39 & 2000  & 11778 & no    & 0.76      & 0.20  & 3.7       \\
    G42 & 2000  & 11779 & no    & 0.76      & 0.26  & 2.9       \\
    G41 & 2000  & 11785 & no    & 0.76      & 0.25  & 3.1       \\
    G55 & 5000  & 12498 & no    & 2.26      & 0.21  & 10.8      \\
    G62 & 7000  & 14000 & yes   & 2.73      & 0.16  & 17.5      \\
    G65 & 8000  & 16000 & yes   & 3.27      & 0.16  & 21.0      \\
    G61 & 7000  & 17148 & no    & 4.36      & 0.21  & 20.9      \\
    G66 & 9000  & 18000 & yes   & 4.19      & 0.16  & 27.0      \\
    G9  & 800   & 19176 & no    & 1.27      & 0.17  & 7.6       \\
    G31 & 2000  & 19990 & no    & 1.16      & 0.16  & 7.4       \\
    G67 & 10000 & 20000 & yes   & 5.06      & 0.16  & 32.5      \\
    G77 & 14000 & 28000 & yes   & 9.64      & 0.16  & 61.8      \\
    G59 & 5000  & 29570 & no    & 4.07      & 0.34  & 11.9      \\
    G81 & 20000 & 40000 & yes   & 19.89     & 0.16  & 125.6     \\
    G64 & 7000  & 41459 & no    & 7.97      & 0.39  & 20.6      \\
    C1  & 10000 & 100E3 & no    & 26.63     & 0.18  & 147.9     \\
    C2  & 10000 & 250E3 & no    & 59.81     & 0.38  & 157.39    \\
    C3  & 10000 & 500E3 & no    & 121.5     & 0.61  & 199.5     \\
    C4  & 10000 & 750E3 & no    & 179.87    & 0.91  & 197.65    \\
    C5  & 10000 & 1E6   & no    & 234.86    & 1.18  & 198.69    \\
    C6  & 100E3 & 5E6   & no    & 1.56E4    & 7.11  & 2200.81   \\
    C7  & 100E3 & 7E6   & no    & 2.05E4    & 9.99  & 2254.74   \\
    \hline
\end{tabular}

    \caption{Performance results comparing GPU performance against CPU performance for the G-set benchmark (G prefix) problems and large custom problems (C prefix).}
    \label{tb:speedup}
\end{table}

\begin{figure}
  \centering
  \includegraphics[width=\columnwidth]{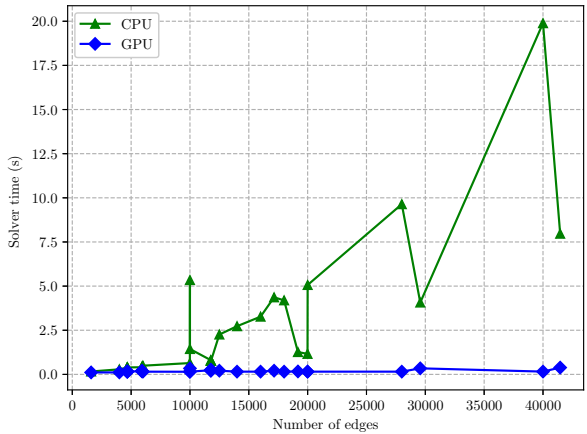}
  \caption{Graphic of the performance in seconds of the GPU and CPU for increasingly large graphs.}
  \label{fig:cpu_v_gpu}
\end{figure}

In Table~\ref{tb:speedup}, the time (in seconds) to complete 1000
simulation sweeps is shown for the CPU and GPU for various G-set
problems of increasing edge counts. In this table, column {\it torus}
indicates whether the graph is a torus or not. Columns $t_{CPU}(s)$ and
$t_{GPU}(s)$ are the run times for CPU and GPU based solutions
respectively. Column {\it speedup} is the speedup of GPU solution over
CPU solution defined as $speedup = t_{CPU}(s)/t_{GPU}(s)$.
Furthermore, we include several very large custom made and randomly
generated non-torus graphs in the table to show the scalability of the
proposed method against the CPU-based solution. It should be noted
that accuracy results are not included for the custom graphs as there
is no data for best known or optimal maximized cut values. Additionally,
Fig.~\ref{fig:cpu_v_gpu} graphically shows the speed results in
seconds for increasing problem sizes and Fig.~\ref{fig:cpu_v_gpu_bars}
shows a bar graph of the results in log scale (the large custom graphs
are omitted from the figures). For all simulations, small and large,
the GPU outperformed the CPU competition. The smaller graph problems
are unable to make best use of the GPU resources, as such, the CPU
performance does not lag far behind. For the very large custom
graphs, the scalability of the GPU can really be seen as the CPU
struggles to handle such large problems while the GPU is able to
finish them quite quickly and even achieve over 2000X speedup for the
largest problems.

 The CPU based Ising solution time grew approximately quadratically with
  the problem size which is expected and is similar to the
  observations of the reported quantum adiabatic computing on the
  exact cover problem~\cite{Farhi:Science'2001}. We can
  also see that the GPU solver time remains quite constant until it
  starts working on the very large random graphs. This nearly constant
  computation time trend can be explained by the GPU architecture. The
  GPU achieves its performance by utilizing a vast number of
  computational resources which enables high throughput
  computations. Because of this, we do not expect the computation time
  to rise dramatically with a rise in problem size so long as the
  computational resources of the GPU are sufficient to handle the
  problem size. In contrast, the CPU's computation time will be
  impacted regardless of the problem size. 


For the most part, as the problem size increased, so did the GPU
speed-up. However, we note that there were some graphs which the CPU
performed quite well and the speedup that the GPU achieved was much
smaller than on other graphs. This is primarily seen in the smaller
graphs where the large amount of compute resources cannot be fully
utilized on the GPU. It is important to note that the GPU
  performance is achievable due to the large amount of compute
  resources that can be utilized in parallel. However, if the size of
  the graph is much larger than the amount of compute resources, then
  we would expect the performance to degrade but also to still be
  better than the CPU. Additionally, for very small graphs that don't
  occupy the GPU resources, we should expect the performance gain
  compared to a CPU based version to be much smaller after further
inspection we identified that the solvers had interesting performance
depending on the graph structure which prompted further investigation.

\begin{figure}
  \centering
  \includegraphics[width=\columnwidth]{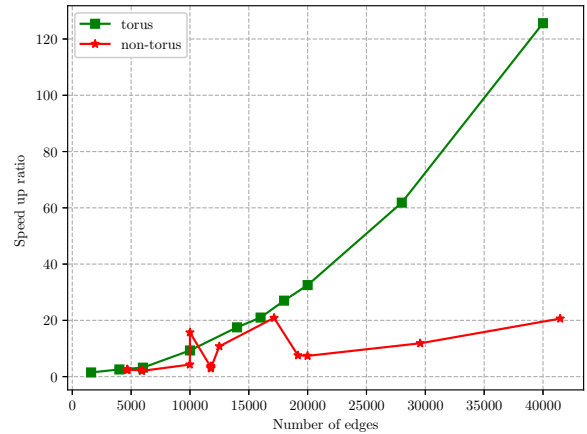}
  \caption{The performance results for the GPU against the CPU, separated by the graph type, torus and non-torus.}
  \label{fig:speedup_torus_v_nontorus}
\end{figure}

We immediately noticed that the performance of both the GPU and
CPU was dependent on whether or not the graph was a 2D torus
structure. As seen in Fig.~\ref{fig:speedup_torus_v_nontorus}, the
speedup (GPU solver time over CPU solver time) is separated by the
graph type, green for a torus and red for a non-torus. By examining
this figure, we see that the speedup of the torus structure, which is
highly regular, steadily increases as the problem size
increases. However, the non-torus structure, while still showing
speedup, is less consistent but generally shows an increase in speedup
as the problem size increases.

\begin{figure}
  \centering
  \includegraphics[width=\columnwidth]{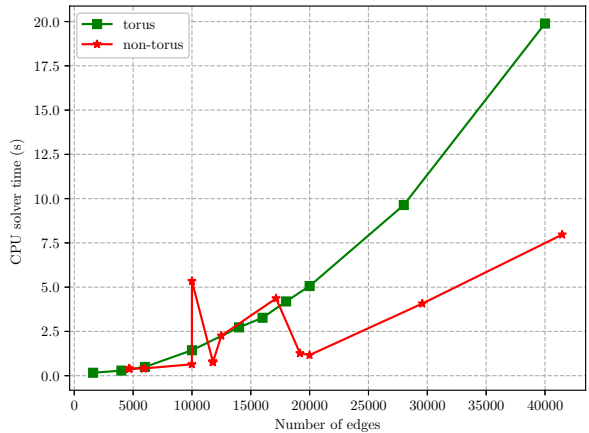}
  \caption{CPU performance comparing torus and non-torus graphs.}
  \label{fig:cpu_torus_v_nontorus}
\end{figure}

\begin{figure}
  \centering
  \includegraphics[width=\columnwidth]{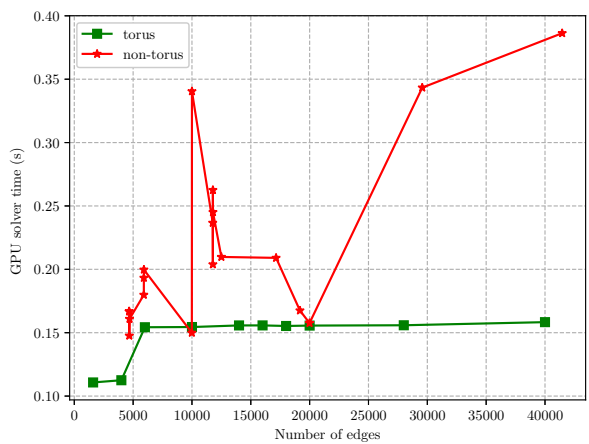}
  \caption{GPU performance comparing torus and non-torus graphs.}
  \label{fig:gpu_torus_v_nontorus}
\end{figure}

We further investigate the individual performance of the CPU and GPU
by plotting their speed results individually and also separating these
results by graph type in Fig.~\ref{fig:cpu_torus_v_nontorus} and
Fig.~\ref{fig:gpu_torus_v_nontorus} respectively. From these graphs we
can make a few observations.For both the CPU and GPU versions, we
  firstly notice that the speed trends when solving the torus
  structures is highly consistent with the increasing problem sizes
  while the non-torus structures have erratic behavior. This can be
  easily explained by the irregularity of the graph structure. Each
  node will have different numbers of edges in the non-torus
  structure. Coincidentally, on the CPU, the torus structures take
  longer to solve than the non-torus structures due to the fact that
  the torus graphs in the g-set benchmark happen to have many more
  nodes than the non-torus structures(see~\ref{tb:speedup} for node
  numbers). More interesting, we see that the GPU performance on the
torus structures is highly efficient with almost no noticeable change
in computation time between the graph with 6000 edges and the graph
with over 40000 edges. This can be explained by the regularity of the
graph structures. This regularity will lead to load balance amongst
the GPU threads during computation which allows the GPU performance to
really shine. We should expect this nearly constant compute complexity
so long as the GPU has sufficient compute resources for the problem
size. Any growth in computation time then, could be attributed to host
to device memory transfers. However, as soon as the resources become
insufficient, certain spin updates will need to be serialized, thus,
decreasing performance.


\section{Conclusion}
\label{sec:concl}
In this work, we have proposed the Ising spin model based computing to
solve the max-cut combinatorial optimization problem, which is widely
used method for VLSI physical design, on the general purpose GPU
(GPGPU).  Our new algorithm is based on the observation that Ising
computing by the simulated annealing process is very amenable to
fine-grain GPU based parallel computing.  GPU-based Ising computing
provides clear advantage as a general solver over existing
hardware-based Ising computing methods that utilize integrated
circuits and FPGAs. We further illustrate how the natural randomness
of GPU thread scheduling can be exploited to during the annealing
process to improve GPU resource utilization.  We also showed that
GPU-based computing can handle any general Ising graph with arbitrary
connections, which was shown to be difficult for FPGA and other
hardware based implementation methods. Numerical results show that the
proposed GPU Ising max-cut solver can deliver over 2000X speedup over
the CPU version of the algorithms over some large examples, which
renders this method very appealing for many practical VLSI physical
design problems.

\label{sec:conclusion}


\bibliographystyle{ieeetr}
\bibliography{../../../bib/physical,../../../bib/security,../../../bib/emergingtech,../../../bib/thermal_power,../../../bib/mscad_pub,../../../bib/interconnect,../../../bib/stochastic,../../../bib/simulation,../../../bib/modeling,../../../bib/reduction,../../../bib/misc,../../../bib/architecture,../../../bib/reliability}

\end{document}